\documentclass{article}
\usepackage[utf8]{inputenc}
\usepackage{color}
\usepackage{url}
\usepackage{graphicx}

\title{Data for Refugees: The D4R Challenge on Mobility of Syrian Refugees in Turkey}
\author{Albert Ali Salah, Alex Pentland, Bruno Lepri, Emmanuel Letouzé, \\ Patrick Vinck, Yves-Alexandre de Montjoye, Xiaowen Dong, \\Özge Dağdelen}
\date{July 2018}

\begin{document}

\maketitle
\begin{abstract}
    The Data for Refugees (D4R) Challenge is a non-profit challenge initiated to improve the conditions of the Syrian refugees in Turkey by providing a special database to scientific community for enabling research on urgent problems concerning refugees, including health, education, unemployment, safety, and social integration.The collected database is based on anonymised mobile Call Detail Record (CDR) of phone calls and SMS messages of Türk Telekom customers. It indicates broad activity and mobility patterns of refugees and citizens in Turkey for one year. The data collection period is from 1 January 2017 to 31 December 2017. The project is initiated by Türk Telekom, in partnership with the Turkish Academic and Research Council (TÜBİTAK) and Boğaziçi University, and in collaboration with several academic and non governmental organizations, including UNHCR Turkey, UNICEF, and International Organization for Migration.
\end{abstract}

\section{Introduction}
After the Syrian Civil War started in 2011-12, civilians in increasing numbers sought refuge in neighboring countries. By May 2017, Turkey had received over 3 million refugees — the largest refugee population in the world. About 30\% of them live in government-run camps near the Syrian border. Many have moved to cities looking for work and better living conditions. They face problems of integration, income, welfare, employment, health, education, language, social tension, and discrimination.\footnote{This is the preprint of the D4R Challenge description paper. Please check the Challenge website for the final version, and the correct citation. Author affiliations: AAS (corresponding author, salah@boun.edu.tr) - Boğaziçi University and Nagoya University; AP - MIT; BL - Fondazione Bruno Kessler; EL - MIT and DataPop Alliance; PV - Harvard University; Y-AdM - Imperial College London; XD - University of Oxford; ÖD - Türk Telekom} 
The Data for Refugees (D4R) Challenge\footnote{\url{http://d4r.turktelekom.com.tr}} is a non-profit project to ultimately improve the conditions of the Syrian refugees in Turkey by providing a special database to scientific community for enabling research on some urgent problems. The challenge datasets opened to the community are based on anonymised mobile Call Detail Record (CDR) of phone calls and SMS messages of Türk Telekom customers. It indicates broad activity and mobility patterns in Turkey for one year. The D4R Challenge, called the Challenge hereafter, opens the data to research groups submitting project proposals, after an evaluation, and on strictly regulated terms. The five focus themes of the challenge are health, education, unemployment, safety, and social integration, respectively. The project is initiated by Türk Telekom, in partnership with the Turkish Academic and Research Council (TÜBİTAK) and Boğaziçi University and in collaboration with several academic and non governmental organizations, including UNHCR Turkey, UNICEF, and International Organization for Migration.

A scientific committee of international experts guides its execution. A Project Evaluation Committee (PEC)\footnote{\url{http://d4r.turktelekom.com.tr/presentation/project-evaluation-committee}} is formed with representatives from academia, government (i.e. ministries related to the challenge), and NGOs working in this area. This committee represents refugee interests, and its job is to ensure that the submitted research projects that are granted access to the data have clear goals, with foreseen benefits to the refugee population in Turkey and elsewhere. Access to D4R data requires PEC approval.

The general aims of the Data for Refugees (D4R) project are to:
\begin{enumerate}
    \item Contribute to the welfare of the refugee populations,
    \item Gain insights on key issues, including safety and security, health, education, unemployment, social integration and segregation, mobility, and distribution of resources and infrastructure,
    \item Help governments and international bodies model the dynamics of the refugee populations and to discover vulnerabilities (socio-economic vulnerabilities, gaps in education and services, etc.),
    \item Seed further projects, co-created with refugees, resulting in new applications, services, and innovative solutions for refugees in Turkey and elsewhere.
\end{enumerate}

The lack of data on refugee mobility is a very important hurdle to the proper functioning of government services and international aid bodies. Innovative approaches attempted to overcome this included the use of satellite imagery to obtain information from the regions in crisis, with limited success~\cite{machado2015analyzing}. This project will, for the first time, allow the analysis of a large-scale mobile CDR database on refugees. The D4D Challenges~\cite{blondel2012data,de2014d4d} have illustrated the usefulness of such data, and the numerous projects completed on these challenges provided insights for social data science. We believe the challenge will be instrumental in understanding refugee mobility, optimizing infrastructure and humanitarian help, for better understanding the effects of interventions in health, education, and integration. Additionally the Challenge, by involving research groups from all around the globe, aims to raise awareness for the refugee issues on a grand scale.

The possibilities that mobile CDR data affords for analysis of a broad set of problems are surveyed in~\cite{blondel2015survey}. Examples of projects conducted with similar data include analysis of disaster resilience~\cite{tomaszewski2014geographic}, infrastructure planning~\cite{martinez2015using}, quantifying mobility effects on the spread of infectious diseases~\cite{baldo2013disease,mari2017big}, developing agent based models for disease migrations~\cite{tompkins2016migration}, disease containment~\cite{lima2015disease}, analysis of community structures and socio-demographic indicators~\cite{trestian2017towards}, detection of unusual events~\cite{gundogdu2016countrywide}, poverty analysis~\cite{pokhriyal2017combining}, mobility during holidays and religious festivals~\cite{scharff2015human}, to name a few. 

The D4R Challenge has a distinct focus around refugee problems, and aims to enable evaluation of refugee related interventions and activities, including, but not limited to, educational activities, social gatherings, NGO actions, government infrastructure investments, etc. It also has the potential to bring insights to the analysis of residential segregation, population structures for specific geographical locations, and factors on social integration~\cite{silm2014ethnic}.

\section{Description of D4R Data}
\subsection{Brief summary}
The D4R Challenge is based on the successful Data for Development (D4D) Challenge series~\cite{blondel2012data}. Three datasets are made available to the challenge participants, along with external helper files. The main difference from D4D is that D4R has a ``Refugee" flag, which indicates (with a high probability) that the CDR belongs to a refugee customer. This flag is given to customers in the database that 1) have ID numbers given to refugees and foreigners in Turkey, 2) registered with Syrian passports, 3) use special tariffs reserved for refugees. None of these groups are guaranteed to include only and exclusively refugees, which serves as a layer of protection: It is not possible to say with certainty that a particular CDR belongs to a refugee or not, but it is only possible to derive patterns from aggregated records. Initially, we have planned to perturb these labels for adding such a protection, but the collection is already noisy, and no further noise is introduced. We list the datasets contained in the Challenge in individual subsections.

We acknowledge that the term ``refugee" is used as a blanket term in the dataset, and includes migrants, asylum seekers, and even foreigners who have acquired a ``temporarily protected foreign individual" ID number in Turkey (i.e. starting with 98 or 99). The dataset needs to be approached with these reservations in mind, and the analysis should carefully consider such biases in the data.

The D4R dataset is collected from 992.457 customers of Türk Telekom, of which 184.949 are tagged as ``refugees'', and 807.508 as Turkish citizens. A total of 1.211.839 subscriptions are included. Of these, 980.697 belong to Turkish citizens, and 231.142 belong to refugees (we refer to these customers as refugees, but as mentioned before, there is some noise in this indicator). Some of the customers had multiple phone lines; each line corresponds to a single subscription. 

75\% of the refugee-tagged customers are recorded as ``male", and 25\% as ``female". There is clearly a gender bias in the ownership of the phone line. This does not mean that 75\% of the phone lines are used by men, however. We have sampled the 807K Turkish customers with the same gender distribution. 

45\% of the refugee customers were registered in Istanbul. Other major cities with refugee presence were Gaziantep, İzmir, Şanlıurfa, and Mersin. We have sampled the Turkish citizen customers mainly from the cities with registered refugee presence, to simplify comparisons.  Table~\ref{table:refugee_distrib} and Table~\ref{table:tc_distrib} show the distribution of customers and their registered cities. Only the top locations are shown. The distribution over all the cities of the country is provided to the participants in the file ``Refugees per city (March 2017).xls". This file shows the official number of refugees registered per city, the official city population in 2017 (excluding refugees), and the percentage of refugees with respect to the city population. Additionally, it shows the number of Türk Telekom customers used for the entire D4R data collection per city, broken into ``refugee" and ``citizen" counts.
The numbers of registered refugees and asylum seekers in Turkey according to registration dates can be obtained from UNHCR websites\footnote{See \url{https://data2.unhcr.org/en/situations/syria/location/113}}. Another useful source of data is the TUIK census estimates for Turkish cities, according to years\footnote{See \url{http://www.tuik.gov.tr/PreIstatistikTablo.do?istab_id=1590}}. This source indicates the population size and growth of each city between 2000-2017.

\begin{table}[tb]
    \centering
    \begin{tabular}{|c|c|c|}
         \hline
         Location & Number of Customers & Percentage \\ \hline
İstanbul & 84.173 & 45,511 \\
Gaziantep &14.898 & 8,055\\
İzmir & 10.425 & 5,637\\
Şanlıurfa & 9.701 & 5,245\\
Mersin & 9.660 & 5,223\\
Hatay & 7.024 & 3,798\\
Ankara & 5.580 & 3,017\\
Konya & 4.718 & 2,551\\
Bursa & 3.479 & 1,881\\
Outside Turkey & 2.902 & 1,569\\
Other & 32.440 & 17,540 \\
\hline
    \end{tabular}
    \caption{The distribution of customers tagged as refugees and their registered locations. Numbers rounded to the third significant digit.}
    \label{table:refugee_distrib}
\end{table}

\begin{table}[tb]
    \centering
    \begin{tabular}{|c|c|c|}
         \hline
         Location & Number of Customers & Percentage \\ \hline
İstanbul & 363.334 & 44,994 \\
Gaziantep &80.655 & 9,988\\
İzmir & 40.501 & 5,016\\
Ankara & 40.443 & 5,009\\
Adana & 40.415 & 5,005\\
Hatay & 40.394 & 5,002\\
Konya & 40.388 & 5,002\\
Antalya &40.367 & 4,999\\
Bursa &40.359& 4,998\\
Şanlıurfa & 40.321 & 4,993\\
Mersin & 40.242 & 4,983\\
\hline
    \end{tabular}
    \caption{The distribution of customers tagged as Turkish citizens and their registered locations. Numbers rounded to the third significant digit.}
    \label{table:tc_distrib}
\end{table}

The usage of the D4R data requires caution in interpreting the representativeness of the data for the refugee population in Turkey. At the end of March 2017, there were 75.724.413 mobile customers in Turkey across all operators (\%94,9 penetration rate)~\cite{btk17}. Excluding machine to machine (M2M) and population of 0-9 age range, the mobile penetration was \%107. According to data from the first three months of 2017, the mobile market share of Türk Telekom (Avea), from which the challenge data is collected, was \%24,7~\cite{btk17}. 

\begin{figure}[h]
\centering
\includegraphics[width=10cm]{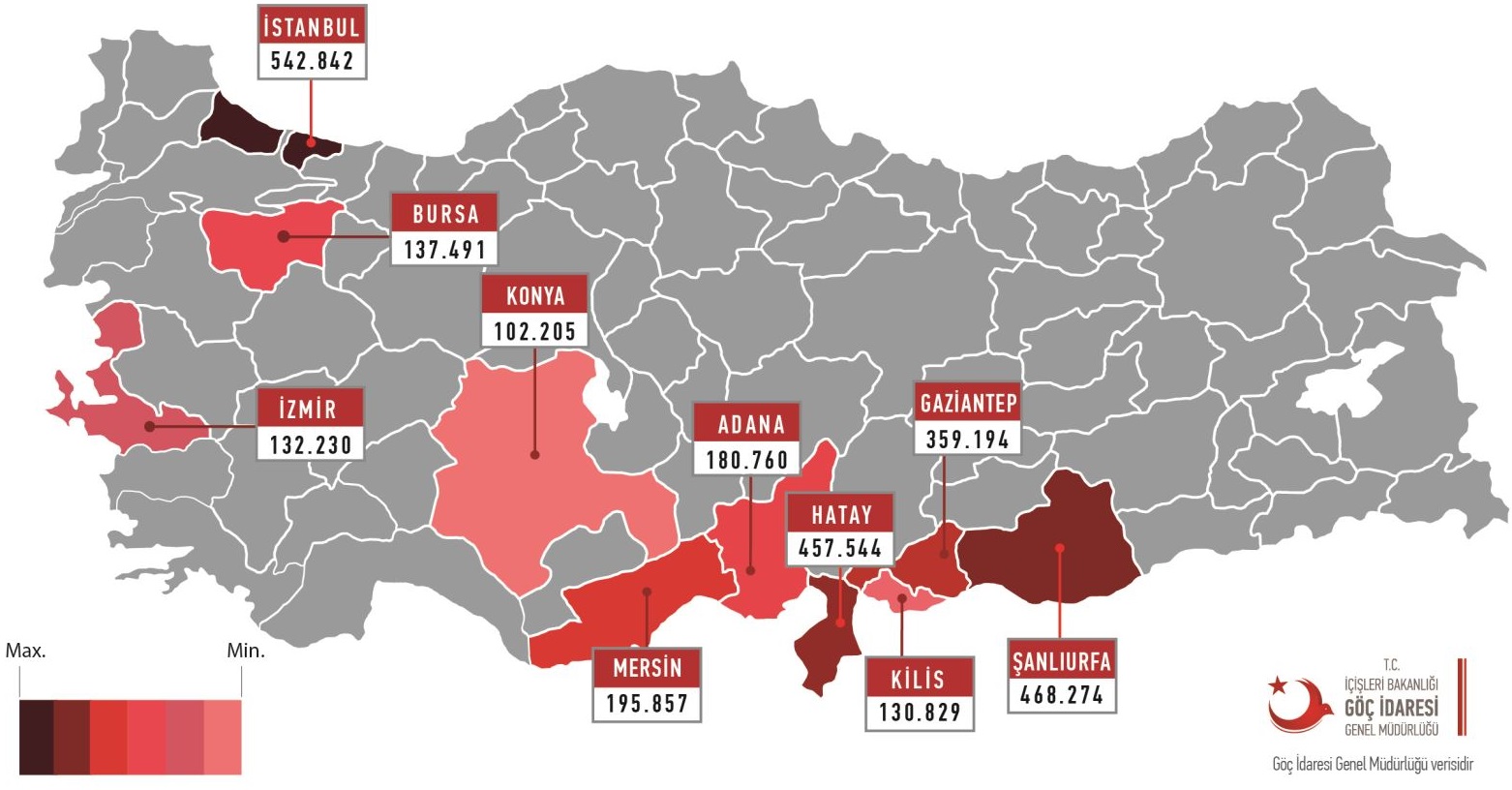}
\caption{The distribution of refugees in the country according to data from Ministry of Interior, Directorate General of Migration Management.}
\label{fig:distribution}
\end{figure}

We have used the entire refugee customer base (with the filtering conditions described before), but the market share of Türk Telekom also shows fluctuations according to the individual cities. Therefore, it is useful to look at official numbers of refugees distributed over the country. This is partly depicted in Fig.~\ref{fig:distribution} as per Ministry of Interior, Directorate General of Migration Management figures\footnote{Only the top ten cities are shown, more detailed information can be obtained from \url{http://www.goc.gov.tr/icerik/migration-statistics_915_1024}.}. However, these figures are from 2018. We have supplied the official figures from March 2017, as mentioned before.

We now describe the contents of the dataset in more detail. 

\subsection{Cell tower locations}
The cell tower (i.e. base station) locations are provided in the file ``Base\_Station\_Location.txt". The file contains the following fields:
\begin{itemize}
    \item BTS\_ID: The identifier of the base tower.
    \item MX\_LAT1,MX\_LAT2,MX\_LAT3: DMS latitude of the base tower.
    \item MX\_LONG1,MX\_LONG2,MX\_LONG3: DMS longitude of the base tower.
    \item MX\_SAHAIL: The registered city of the base tower.
    \item MX\_SAHAILCE: The registered district of the base tower.
    \item MX\_POPAREA: An inofficial note about the population type around the base tower, used internally in Türk Telekom. It takes values of RURAL, SUB\_URBAN, INDUSTRIAL, SEASONAL AREAS, DENSE\_URBAN, HOT SPOT, OPEN IN URBAN, AIRPORT, SUBURBANLOW, POPRURAL, and INDOOR.
\end{itemize}
In some rare cases, the precise location information of the base station is missing, only the city is indicated.
The interpretation of the latitude and longitude follows degree, minutes, seconds (DMS) syntax. For example, the district of BARTIN in the city of ZONGULDAK is represented by these six numbers as follows: (41 25 43.1184 32 4 37.9344). This corresponds to $41^\circ 25' 43.1184''$ N DMS latitude, and $32^\circ 4' 37.9344''$ E longitude.
\subsection{District locations}
To disambiguate the base stations, we provide a file that contains district coordinates. This file, named ``district\_coordinates.csv", has the following fields:
\begin{itemize}
    \item CITY: Name of the city.
    \item DISTRICT: Name of the district.
    \item POPULATION\_2014: The official census population of the district in 2014.
    \item LATITUDE, LONGITUDE: The 2D coordinates of the district.
    \item MX\_LAT1, MX\_LAT2, MX\_LAT3, MX\_LONG1, MX\_LONG2, MX\_LONG3: The DMS coordinates of the district.
\end{itemize}

A conversion script (such as \url{https://www.latlong.net/lat-long-dms.html}) can be used to convert the (latitude, longitude) variables into DMS. For example, the 2D coordinates of (41.428644	32.077204) for Zonguldak, Bartın, translate into the DMS coordinates given in the previous subsection.
\subsection{Dataset 1: Antenna Traffic}
The first database we provide includes one year site-to-site traffic on an hourly basis. This dataset contains the traffic between each site for a year. Calls between Türk Telekom (TT) customers and other service providers (SP) only have information about the TT side. For each record, total number and duration of calls are recorded in an aggregated fashion. 

The database is split into voice and SMS partitions. For the voice partition, the file ``Dataset 1\_2017XX.txt'' contains the data for month XX, and there are 12 such files. Each file contains the following fields:
\begin{itemize}
\item TIMESTAMP: Day and hour considered in format DD-MM-YYYY HH (24 hours format).
\item OUTGOING\_SITE\_ID: ID of the site the call originated from. Unknown stations are denoted as ``-99'' or ``9999''.
\item INCOMING\_SITE\_ID: ID of the site receiving the call.
\item NUMBER\_OF\_CALLS: The number of calls in this 1 hour slot.
\item NUMBER\_OF\_REFUGEE\_CALLS: The number of calls involving numbers tagged as ``refugee''.
\item NUMBER\_OF\_TOTAL\_CALL\_DURATION: The total call duration from all calls.
\item REFUGEE\_CALL\_DURATION: The total call duration from calls involving numbers tagged as ``refugee''.
\end{itemize}

For the SMS partition, the file ``Dataset 1\_SMS\_2017XX.txt'' contains the data for month XX, and there are 12 such files. Each file contains the following fields:
\begin{itemize}
\item TIMESTAMP: Day and hour considered in format DD-MM-YYYY HH (24 hours format).
\item OUTGOING\_SITE\_ID: ID of site the SMS originated from. Unknown stations are denoted as ``-99'' or ``9999''.
\item INCOMING\_SITE\_ID: ID of the site receiving the SMS.
\item NUMBER\_OF\_SMS: The number of SMS messages in this 1 hour slot.
\item NUMBER\_OF\_REFUGEE\_SMS: The number of SMS messages involving numbers tagged as ``refugee''.
\end{itemize}
\subsection{Dataset 2: Fine Grained Mobility}
The dataset contains cell tower identifiers used by a group of randomly chosen active users to make phone calls and send texts. The data are timestamped and a particular group of users are observed for a period of 2 weeks. At the end of the two-week period, a fresh sample of active users are drawn at random. We provide data for the entire 1-year sampling period. The users are represented by random numbers in the dataset, and no personal information is stored. These numbers start with 1 for refugees, 2 for non-refugees, 3 for unknown. However, this indicator should be considered to be somewhat noisy. Among the users who are marked as refugees, there may be customers who are not refugees, and vice versa. Consequently, it will not be possible to say with 100\% certainty whether an invitation CDR belongs to a refugee or not. (This numbering scheme is also used for Dataset 3.) There is no identifying information about the other party of the call; only the callee prefix (1: refugee, 2: not refugee, 3: unknown) is given. 

To protect privacy, new random identifiers are chosen for every two-week period, and if a user is sampled in more than one period, these records cannot be associated with each other. For missing antenna locations, a code of -99 or 9999 is assigned. This dataset is also separated into voice and SMS partitions. Furthermore, to deal with large file sizes, it is further divided into files containing incoming (IN) and outgoing (OUT) calls, resulting in four files per 15-day period.

The files ``Dataset 2\_2017XXW\_In.txt," ``Da-taset 2\_2017XXW\_Out.txt," ``Dataset 2\_2017XXW\_SMS\_In.txt," and ``Dataset 2\_2017XXW\_SMS\_Out.txt" all have similar structure, where XX ranges from 01 to 26, and denotes a 15-day period, starting from 1-15 January 2017 in the first file. The files have the following fields: 
\begin{itemize}
    \item CALLER\_ID: Random number for a user generated specifically for this 15-day period. Note that the user is not necessarily the initiator of the call, that is determined by the CALL\_TYPE flag. The first digit denotes ``refugee'' (1), ``non-refugee'' (2), ``unknown'' (3). 
    \item TIMESTAMP: Day and hour considered in format DD-MM-YYYY HH (24 hours format).
    \item CALLEE\_PREFIX: A value for the other party of the call that denotes ``refugee'' (1), ``non-refugee'' (2), ``unknown'' (3).
    \item SITE\_ID: The ID of the base station.
    \item CALL\_TYPE: The Call type is either outgoing (1) or incoming (2).
\end{itemize}

\subsection{Dataset 3: Coarse Grained Mobility}
In this dataset, the trajectories of a randomly selected subset of users are provided for the entire observation period, but with reduced spatial resolution. We divide the entire country into the electoral prefectures (or districts), and for each call record, only the prefecture information is provided. The IDs are randomly assigned, and two different users may have the same ID in Dataset 2 and Dataset 3. The database is split into incoming (IN) and outgoing (OUT) calls to deal with large files. The files ``Dataset 3\_2017XX\_In.txt" and ``Dataset 3\_2017XX\_Out.txt" have a similar structure, and contain the fields:
\begin{itemize}
    \item CALLER\_ID: The randomly assigned ID of the user (different from Dataset 2). Similarly to Dataset 2, the call initiator is determined by CALL\_TYPE.
    \item TIMESTAMP: Day and time considered in format DD-MM-YYYY HH:MM (24 hours format).
    \item ID: The ID of the district.
    \item CITY\_ID: The ID of the city.
\end{itemize}

In order to achieve the mapping to the cities and prefectures, two additional files are provided. In ``Dataset 3\_City Mapping.txt,'' the CITY\_ID is followed by CITY\_DESC, which is the name of the city. 81 cities are included. In ``Dataset 3\_District Mapping.txt,'' the ID field represents the district ID, as used in the Dataset 3, and the BTS\_DISTRICT field gives the name of the district. There are a total of 1025 districts.
\section{Ethical and privacy issues}
In this section, we briefly discuss the ethical and privacy issues regarding the Challenge data. The collection, storage and protection of data in the D4R Challenge complies with European Union requirements regarding the protection of personal data and the protection of privacy in the electronic communications sector. Furthermore, research on the previously conducted D4D Challenges established that the data such as offered in this project does not allow identification of individuals~\cite{al2015literature,cecaj2016re,gambs2014anonymization,taylor2016no}. Sharad and Danezis note that providing aggregated data such as antenna traffic results in ``little scope of privacy breach\ldots since it contains no personally identifiable information about the users. It could be used to study traffic patterns during the entire period but reveals no information pertaining to the users."~\cite{sharad2013anonymizing}.

\textbf{Definitions:} Personal data means any information relating to an identified or identifiable natural person. Personal data does not include anonymous information, that is, information that does not relate to an identified or identifiable natural person or to data rendered anonymous in such a way that the Data Subject is not or is no longer identifiable. Data Subject means a natural person (i.e. an individual) who can be identified directly or indirectly, in particular by reference to Personal Data.

\textbf{Consent, legitimate and fair processing:} The data in the Database comes exclusively from Türk Telekom customers, who has consented to its anonymised use for research purposes through the mandatory user agreement at the time of the purchase of the phone line. The content of phone activity, actual phone numbers, identities, addresses, or similar personal information are neither stored, nor distributed with the Database. Subsequently, it is not possible to identify natural persons with the Database.

\textbf{Transparent processing:} The nature of the data, the assurance of its anonymity, as well as the ethical precautions to ensure its proper use are (at the time of the start of the Challenge) documented openly on the Challenge website. Accessible and plain language is used, and further contact information is supplied to respond to questions about the data usage.

\textbf{Project Evaluation Committee (PEC):} The PEC is formed with representatives from academia, government, and related NGOs. Its aim is to represent refugee interests in the Challenge, and all project proposals are pre-screened by the Scientific Committee and by PEC. The proposals that pass the initial screening are granted access to the dataset, upon submitting the signed User Agreement form.
Criteria for passing the initial screening were:
\begin{itemize}
\item A project proposal is submitted (in English, which is the common language for the international PEC), and all the team members who will access the data are identified.
\item The project PI has a permanent affiliation. 
\item The project uses the D4R data meaningfully. 
\item The project aims are aligned with the goals described in the previous section, do not represent a commercial interest, and do not endanger the privacy or well-being of individuals or groups.
\end{itemize}

Data access for the Database is granted to participants during the designated Challenge period, by a mandatory user agreement prepared by Türk Telekom lawyers, and approved by ADIEK, the Ethical Conduct committee of Boğaziçi University. The agreement permits third parties to analyze the anonymised and aggregated data, summarized previously, to submit a research report at the end of the Challenge and to present the results at a special workshop. A white paper is be prepared to inform the related government bodies and NGOs about the results of the projects. The project reports are published publicly on the project website, after evaluation by the D4R PEC and D4R SC. There are three possible outcomes for submitted reports: (1) Normal publication, for papers that treat the ethical issues correctly. (2) Ask to consider adjustments, for papers that require amendments and removal of sensitive material before publication. We ask the projects to be careful not to include statements that may harm the refugee population in any way, or may promote negative sentiments about the refugee populations. (3) Do Not Publish, either because the report is on a sensitive issue, or because it is not scientifically rigorous and its conclusions are not warranted. Sensitive reports may be shared directly with related institutions or authorities. The PEC and the project proposers jointly decide on this, on a case by case basis. 

\textbf{Retention, destruction and archiving:} The Challenge mandates that all participants destroy the Database upon completing the challenge. Any publication based on the Database requires the pre-approval of the Project Evaluation Committee. Any further use of the Database (for instance to complete numerical experiments for a publication under review) will be regulated by PEC, and extensions are to be conditionally granted for specific purposes on a case by case basis.

\textbf{Information:} Every care has been given to ensure that the information provided in the Database does not cause any harm, prejudice, or distress to customers, regardless of their refugee status. PEC provides an additional layer of control, and will examine project reports confirm to this maxim.

\textbf{Access, correction, erasure, objection:} The Database does not contain personal information, and it is not possible for individuals to request access to personal data. The data are anonymised and aggregated in a way to prevent identification of persons. For the same reason, correction, erasure, and objection do not apply for the Database. We note that this is a stricter protection condition than most envisioned CDR usage scenarios. Furthermore, the Database does not contain children’s data, as each registered customer has to be over 18 years of age, and thus legally permitted to own and use a mobile phone line. Personal profiling (such as used for CRM applications) is not possible with the Database.

\textbf{Responsibility and accountability:} The responsibilities of all parties concerned are defined clearly, and set out in the data agreement prepared by TT lawyers.

\textbf{Data protection by design and default:} Data collection follows this principle, where any names, real phone numbers, or other identifying information is excluded from the design of the Database. The pseudo-random numbers representing customers are not stored anywhere along with actual phone numbers. Subsequently, the anonymisation works only one way. Refugee status is indicated by purposefully noisy indicators, and no effort is spent to ensure its validity. Subsequently, only aggregate-level conclusions can be drawn from the Database. It is not possible to use the Database for the surveillance and tracking of individuals. 

\textbf{Limitations:} Data access is not provided to institutions in a blanket permission, but to specific individuals within institutions, whose names and roles in the proposed research project is clearly defined in the data agreement.
\section{Acknowledgments}
We thank Duhan Can Çakı (Türk Telekom), Salim Yılmaz (Türk Telekom), Oktay Namver (Türk Telekom), Ali Görçin (TÜBİTAK BİLGEM), Merve Astekin (TÜBİTAK BİLGEM), and the members of the D4R PEC for their contributions. This work is partially supported by a MIT MISTI grant to Alex Pentland and Albert Ali Salah.

\end{document}